Universal Chemomechanical Design Rules for Solid-Ion Conductors to Prevent Dendrite Formation in Lithium Metal Batteries

Chengyin Fu[1], Victor Venturi[2], Zeeshan Ahmad[2], Andrew W. Ells[1], Venkatasubramanian Viswanathan[2,3], Brett A. Helms[1,4,*]

[1] The Molecular Foundry, Lawrence Berkeley National Laboratory, 1 Cyclotron Road, Berkeley, CA 94720, USA
[2] Department of Mechanical Engineering, Carnegie Mellon University, Pittsburgh, Pennsylvania 15213, USA
[3] Department of Materials Science and Engineering, Carnegie Mellon University, Pittsburgh, Pennsylvania 15213, USA
[4] Materials Sciences Division, Lawrence Berkeley National Laboratory, 1 Cyclotron Road, Berkeley, CA 94720, USA

Dendrite formation during electrodeposition while charging lithium metal batteries compromises their safety.[1–6] While high shear-modulus ($G_s$) solid-ion conductors (SICs) have been prioritized to resolve pressure-driven instabilities that lead to dendrite propagation and cell shorting, it is unclear whether these or alternatives are needed to guide uniform lithium electrodeposition, which is intrinsically density-driven.[7–9] Here, we show that SICs can be designed within a universal chemomechanical paradigm to access either pressure-driven dendrite-blocking or density-driven dendrite-suppressing properties, but not both. This dichotomy reflects the competing influence of the SIC's mechanical properties and partial molar volume of Li$^+$ ($V_{Li+}$) relative to those of the lithium anode ($G_{Li}$ and $V_{Li}$) on plating outcomes.[9] Within this paradigm, we explore SICs in a previously unrecognized dendrite-suppressing regime that are concomitantly "soft", as is typical of polymer electrolytes, but feature atypically low $V_{Li+}$, more reminiscent of "hard" ceramics. Li plating (1 mA cm$^{-2}$; $T$ = 20 °C) mediated by these SICs is uniform, as revealed using synchrotron hard x-ray microtomography. As a result, cell cycle-life is extended (>300 cycles *vs.* ~100 cycles for control cells), even when assembled with thin Li anodes (~30 μm) and high-voltage NMC-622 cathodes (1.44 mAh cm$^{-2}$), where ~20% of the Li inventory is reversibly cycled.

Heterogeneous nucleation and ramified growth of lithium metal electrodeposits while charging lithium metal batteries is tied to uneven Li$^+$ transport across the anode–electrolyte interface.[7–11] Whereas the increasingly fractal character of this interface during battery cycling accelerates electrolyte degradation, rare events associated with dendrite formation, if left unchecked, can lead to device shorting and in some cases thermal runaway.[1–6] Both early- and late-stage instabilities associated with dendrite formation and propagation can be modeled using Butler–Volmer physics,[8]

$$\frac{i_{deformed}}{i_{undeformed}} = exp\left(\frac{(1-\alpha_a)\Delta\mu_{e^-}}{RT}\right),$$

where $i$ is the current density at either a deformed or undeformed interface, $\alpha_a$ is the anodic charge transfer coefficient, and $\Delta\mu_{e^-}$ is the change in electrochemical potential of the electron at a deformed interface. Notably, $\Delta\mu_{e^-}$ is governed by the Li–SIC interfacial tension ($\gamma$) as well as the deviatoric and hydrostatic stresses in the system, according to,

$$\Delta\mu_{e^-} = -\frac{1}{2z}(V_{Li} + V_{Li^+})\left(-\gamma\kappa + \boldsymbol{e_n} \cdot [(\boldsymbol{\tau_d^{Li}} - \boldsymbol{\tau_d^s})\boldsymbol{e_n}]\right) + \frac{1}{2z}(V_{Li} - V_{Li^+})(\Delta p^{Li} + \Delta p^s),$$

where $\kappa$ is the mean curvature at the interface, $\tau_d^{Li}$ and $\tau_d^s$ are the deviatoric stresses (i.e., the traceless part of the stress tensor) at the Li electrode and electrolyte sides of the interface, and $\Delta p^{Li}$ and $\Delta p^s$ are the associated gage pressures. This model assumes that the sole cause of the change in chemical potentials of the different species involved in Li electrodeposition are the mechanical stresses and interfacial surface tension. To ensure that $i_{deformed}$ is lower than $i_{undeformed}$ at the dendritic tips and higher in the valleys for stable electrodeposition, it has been shown that $\Delta\mu_{e^-}$ should be negative.[7–9] In that $\gamma$ is negligible and the deviatoric stress is always destabilizing, the hydrostatic stress term in this expression dominates $\Delta\mu_{e^-}$ and therefore dictates plating outcomes (Fig. 1a).

The hydrostatic stress term is nominally a function of $V_{Li^+}$, $V_{Li}$, $G_s$, and $G_{Li}$; in addition, the volume ratio $v = V_{Li^+}/V_{Li}$ governs its sign. For $v > 1$, $\Delta\mu_{e^-}$ becomes negative only when $G_s/G_{Li} > 2.2$, which agrees well with the prediction of Monroe and Newman that SICs with $G_s/G_{Li} > 2$ are needed to block dendrite propagation in lithium-ion and lithium metal batteries.[6,7] In such instances, high shear-modulus SICs for which $v > 1$ are characteristically reconfigurable with respect to their ion-conducting domains, but require placement within a suitably rigid host matrix, as might be possible with block co-polymer and related electrolytes.[12–14] However, it follows from our expanded chemomechanical model that high shear-modulus SICs with minimally reconfigurable ion-conducting domains, such as ceramics, are outside the predicted stable plating regime, due to density-driven rather than pressure-driven instabilities. Recent studies of both block copolymer and ceramic SICs in Li metal cells are consistent with our analysis.[15–17]

Somewhat surprisingly, then, $\Delta\mu_{e^-}$ becomes negative when $G_s/G_{Li} < 0.7$ and when $v < 1$, with slight variations linked to the SICs Poisson's ratio (Fig. 1a).[9] In other words, density-driven suppression of dendrites is achievable with "soft" SICs that feature low volume changes as $Li^+$ deposits at the anode–electrolyte interface. This is an unusual test-case for dendrite-suppressing SICs, since known "soft" electrolytes, particularly polymer electrolytes, typically undergo large volume changes as $Li^+$ exits transiently-formed solvation "cages", resulting in $v > 1$. Thus, to access dendrite-suppressing character, SICs should be re-designed with minimally reconfigurable, ceramic-like, ion-conducting domains embedded in a soft, polymer-like matrix with relatively low shear modulus. In that the ion current should be high to avoid excessive cell polarization during cycling, maximizing the interface area between constituents in such a hybrid is also desirable,[18] but rarely encountered or studied systematically within such a chemomechanical paradigm.

Here, we capture both the low $V_{Li^+}$ of inorganic SICs and the low $G_s$ of polymers in a hybrid class of nanostructured SICs that transport $Li^+$ along their heteromaterial interfaces and confirm their predicted dendrite-suppressing character in high-voltage Li metal cells with an N/P ratio of ~4 (i.e., the ratio of anode to cathode capacity). As the ion-conducting inorganic phase, we turned to lithium halides (LiX, where $X^- = F^-$, $Cl^-$, $Br^-$, or $I^-$) whose bulk and interfacial ionic conductivity spans $10^{-6}$–$10^{-2}$ S cm$^{-1}$ at ambient temperature.[19–21] Whereas all are reductively stable on lithium metal, only LiF is oxidatively stable against high-voltage NMC-622 cathodes. LiF is also dimensionally stable in carbonate electrolytes commonly used in Li–NMC-622 cells, due to its high enthalpy of formation and hence low solubility therein. This allows us to make use of them in nanostructured hybrids without morphological evolution. Unfortunately, due to its high enthalpy of formation, LiF is notably difficult to nano-structure in a soft polymer matrix.[22–24]

**Synthesis of Dendrite-Suppressing LiF@PIM SICs by *In-Situ* Cation Metathesis**

We were successful in generating hybrid nano-LiF@polymer hybrid SICs using an *in-situ* cation metathesis. Soluble tetraalkylammonium fluoride precursors (e.g., tetrabutylammonium fluoride, TBAF) to insoluble LiF were loaded (0–60% *w/w*) into a microporous polymer host (e.g., PIM-1)[25,26] and applied as a coating (typically, 0.5–2 µm) from homogeneous inks on a polyolefin separator (e.g., Celgard 2325). The coated separator was then assembled in either Li–Li or Li–NMC-622[27] cells along with a carbonate electrolyte containing an ionizable lithium salt (e.g., $LiPF_6$). The cell was then aged to interconvert the polymer-embedded TBAF to the thermodynamically more favorable LiF using the electrolyte's reservoir of lithium ions (Fig. 1b). The distribution of LiF in the polymer after cation metathesis was homogeneous, as evidenced by energy dispersive spectroscopy (EDS) of the composites (Supplementary Fig. 1). Given the overall simplicity of this scheme, a lithium electrode laminated with our TBAF@PIM-1 coated polyolefin separator constitutes an attractive component for battery manufacturing: a lithium electrode sub-assembly (LESA).

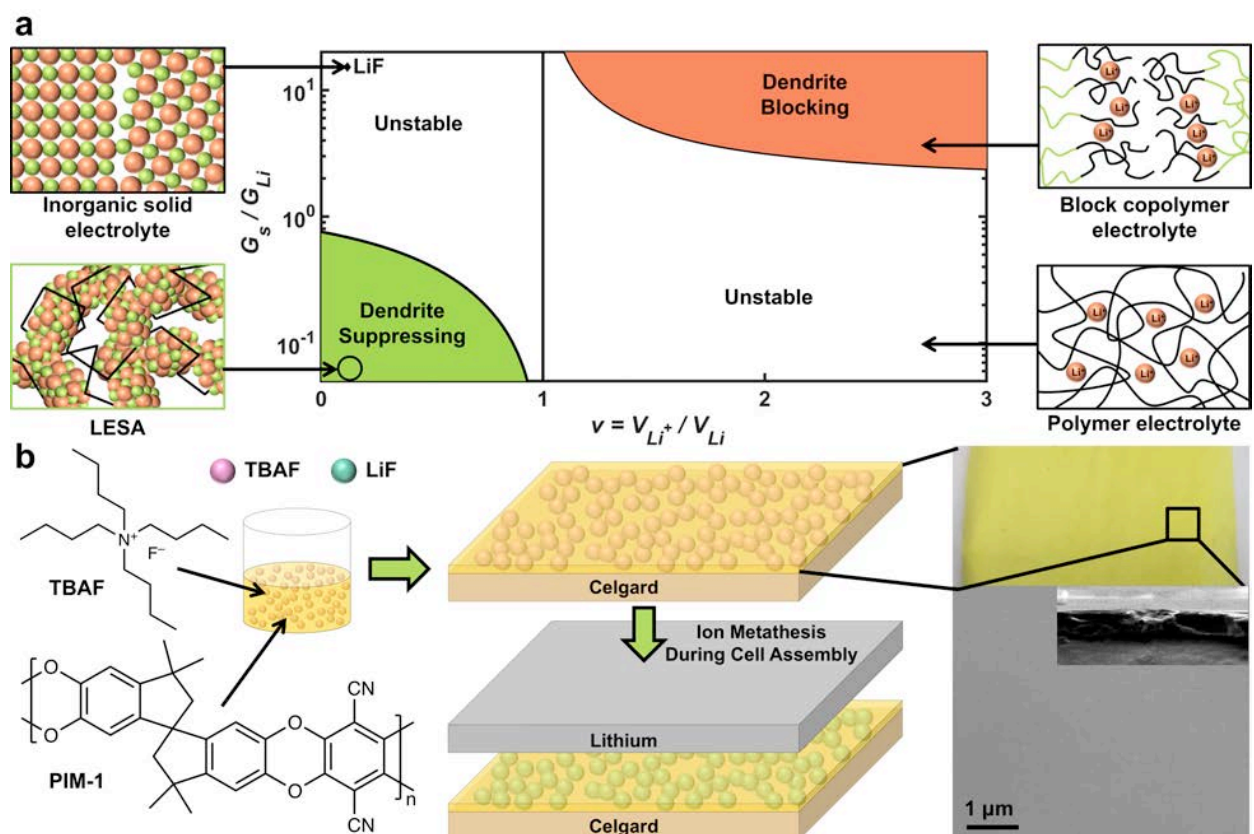

**Fig. 1: Classifying solid-ion conductors within a universal chemomechanical model for dendrite formation during electrodeposition.**
**a,** Chemomechanical model underlying the successes and failures of solid-ion conductors (SICs) in stabilizing lithium metal anodes while batteries incorporating them are charging. SICs can access either dendrite-blocking character or dendrite-suppressing character, but not both. Prototypical SICs that serve as the battery's electrolyte are labeled in each region. This work focuses on SICs that are dendrite-suppressing, on account of their comparably low shear modulus $G_S$ (relative to $G_{Li}$) and low Li$^+$ partial molar volume $V_{Li+}$ (relative to $V_{Li}$). Such SICs are comprised of "soft" polymers infiltrated with nanostructured "hard" ceramics and transport Li$^+$ across their heteromaterial space-charge regions. **b,** Homogeneous inks containing tetrabutylammonium fluoride (TBAF) and microporous polymer PIM-1 are first coated onto a mesoporous polyolefin support prior to integration with the Li anode in Li metal cells. This construct is denoted as a lithium electrode sub-assembly (LESA). Once the cell is assembled, *in-situ* cation metathesis interconverts TBAF in the polymer into nano-LiF using the electrolyte's reservoir of lithium ions.

Cation metathesis results in a volume change that is commensurate with the amount of TBAF initially loaded into the film. We characterized the influence of TBAF loading on the resulting morphologies of LiF@PIM composites using SEM, which showed few aberrations for loadings of 0–5% (*w/w*); here, PIM-1 accommodates the volume change effectively (Fig. 2a–d). At TBAF loadings exceeding ~5% (*w/w*), the composite morphology is increasingly porous (Fig.

2e,f). We characterized the length-scale of LiF domains generated (11.8–13.2 nm) using Scherrer analysis of the X-ray diffraction data (Fig. 2g and Table 1), and the final LiF loading was evaluated *ex-situ* using TGA (Table 1 and Supplementary Fig. 2).

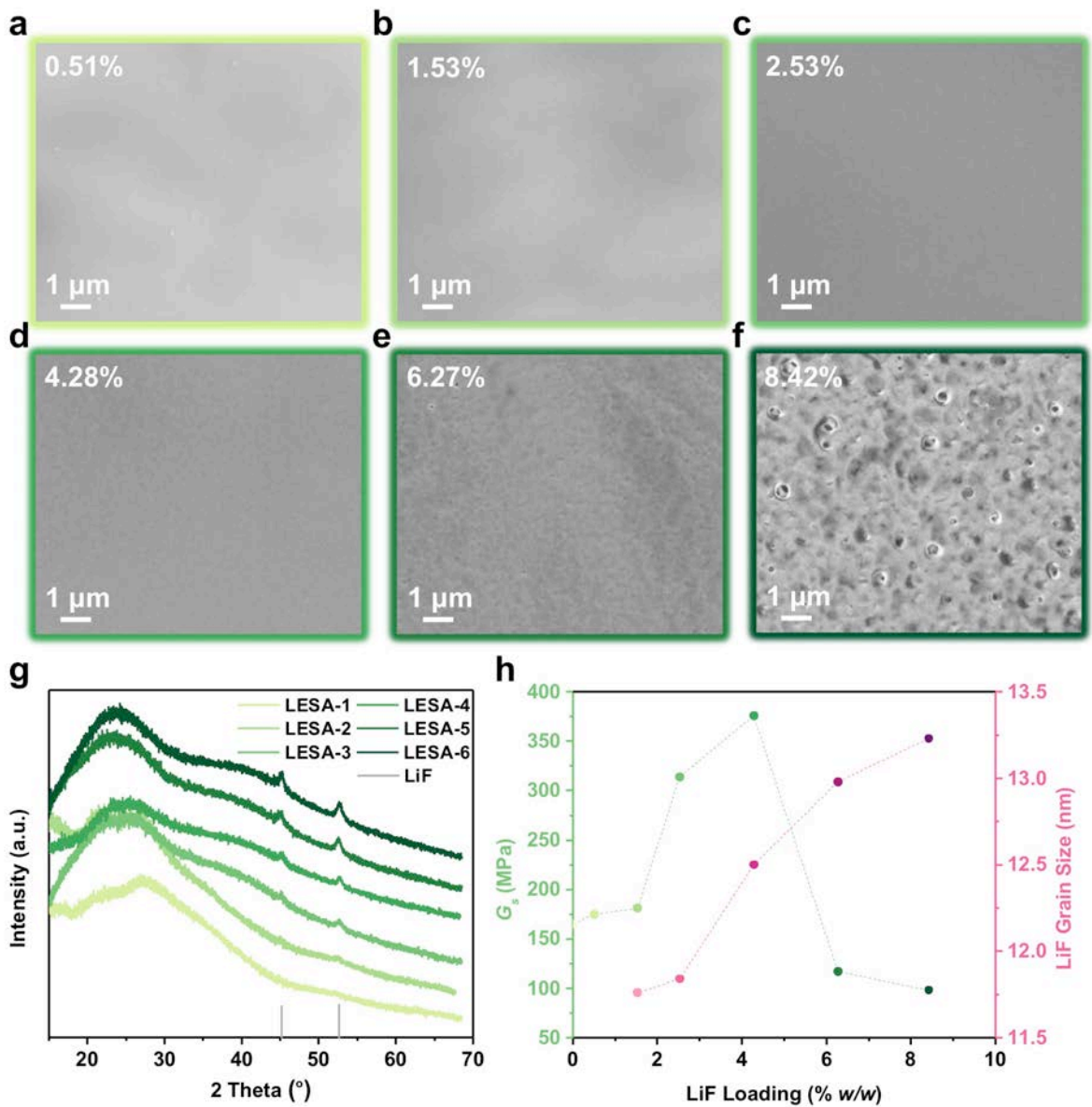

**Fig. 2: Formulation dependent architectures, morphologies, and mechanical properties evidenced for LiF@PIM composites generated *in-situ* by cation metathesis.**
**a–f,** Top-view scanning electron micrographs of LiF@PIM composites, denoted LESAs 1–6, which vary in LiF content, spanning 0.5–8.5% *w/w* (Table 1). **g,** Confirmation that cation metathesis yields nano-LiF@PIM composites was gleaned from XRD of LESAs 1–6; reference peaks for LiF are depicted as grey bars. **h,** Shear modulus, $G_s$, as determined by dynamic mechanical analysis (Supplementary Fig. 3) and upper-bound estimates of LiF domain sizes in LiF@PIM composites, LESAs 1–6, determined by Scherrer analysis of the XRD data shown in **g**.

**Table 1. Chemomechanical characteristics of LiF@PIM nanocomposites, LESAs 1–6.**

|  | PIM-1 | LESA-1 | LESA-2 | LESA-3 | LESA-4 | LESA-5 | LESA-6 |
|---|---|---|---|---|---|---|---|
| LiF grain size (nm) | N/A | N/A* | 11.76 | 11.84 | 12.50 | 12.98 | 13.23 |
| LiF loading (% w/w) | 0 | 0.51 | 1.53 | 2.53 | 4.28 | 6.27 | 8.42 |
| $G_s$ (MPa) | 165 | 175 | 181 | 314 | 376 | 117 | 98 |
| $\sigma_{20C}$** ($10^{-5}$ S cm$^{-1}$) | 0.23 | 0.29 | 0.57 | 1.33 | 1.52 | 1.65 | 3.30 |

\* LESA-1 was not amenable to Scherrer analysis, as LiF peak intensity in XRD was too weak.
\*\* Determined by EIS for cells assembled with PIM-1 or LESAs 1–6 (Supplementary Fig. 4).

**Chemomechanical Characterization of Hybrid LiF@PIM SICs**

Loading nano-sized inorganics in polymers typically increases their tensile strength and shear modulus, which can be evaluated by dynamic mechanical analysis (DMA). From stress–strain curves acquired for macroscopic specimens of each composite (Supplementary Fig. 3), we noted deviations from the expected mechanical behavior as a function of LiF loading (Fig. 2h). In the absence of LiF, PIM-1 has a shear modulus, $G_s$ = 165 MPa. Introducing nano-LiF in PIM-1 monotonically increases $G_s$ to a maximum of 376 MPa when the LiF loading is 4.3% (w/w). However, when the LiF content is increased further, $G_s$ decreases significantly, which we attribute to the emergent porosity of those composites arising from volume changes during the *in-situ* cation metathesis (Fig. 2e,f). Commensurate with these observations, composites with higher LiF loading experience less deformation before breaking. Within this design space, then, we access $G_s/G_{Li}$ values of 0.023–0.09, indicating nanostructured LiF@PIM-1 SICs are "soft" as designed.

The molar volumes of Li and Li$^+$ determine the change in chemical potential $\mu$ of these species due to pressure according to the relation: $\Delta\mu = (\partial\mu/\partial p) \Delta p = V\Delta p$. The molar volume of Li in Li metal can be easily obtained as the inverse of its molar density. In contrast, the partial molar volume of Li$^+$ in the electrolyte, $V_{Li^+}$, is hard to determine by direct measurement. For binary electrolytes, the Newman–Chapman relation states that the partial molar volume of an ion is inversely proportional to its transference number.[28] However, this relation is not applicable for hybrid SICs, including ours. Therefore, we determined $V_{Li^+}$ computationally by investigating the atomistic environment around the Li$^+$ species during ionic conduction at the surface of LiF.

The volume of the ion was determined using Bader charge analysis.[29] An LiF (100) slab with 8% Li vacancy concentration was used as the model structure. Decreasing the Li vacancy concentration resulted in a change of less than 10 meV in the activation energy (Supplementary Fig. 5). Li$^+$ at the surface was simulated hopping between its original site to a vacant site in five stages. The charge density required in the Bader charge analysis was obtained for the different stages of hopping using density functional theory (DFT) calculations. Supplementary Fig. 6 shows isosurfaces of charge density used to calculate $V_{Li^+}$. From the calculation of the Bader volume of Li$^+$, we obtained a value of $v$ equal to 0.21, indicating that we are accessing the density-driven stability regime through our hybrid SIC (Fig. 1a). Bader volume of atoms capture reliably trends in molar volume of atoms in molecules and solids.[30] We note that for LiF@PIM,

the presence of the polymer in the vicinity of the inorganic phase may lead to a small increase in the Bader volume, but we expect this effect to be small due to the absence of highly electronegative species that can bind to the LiF surface and that $v$ will therefore remain less than 1.

It has been previously reported that LiF has a low surface diffusion barrier for Li$^+$,[31] which has been explained to be the cause of uniform lithium electrodeposition and dendrite suppression in Li anodes protected by it.[22,23] However, previous DFT calculations of surface diffusion barrier used the Perdew–Burke–Ernzerhof exchange correlation functional,[45] which has been known to grossly underestimate barrier heights.[32] Here, we used the Bayesian error estimation functional (BEEF-vdW),[33] which includes non-local van der Waals correlation and has been shown to perform better in accuracy than other functionals for the calculation of barrier heights[32]. Using this functional, we carried out nudged elastic band[34] calculations on the LiF (100) surface slab used for Bader charge analysis. The energies of the five stages during Li hopping were calculated using DFT and the activation energy was obtained as the difference between the highest and the lowest energies. The activation energy ($E_a$) obtained was 0.34±0.13 eV. In contrast, the value for bulk LiF using the same method was found to be 0.62±0.17 eV. The uncertainty estimate for $E_a$ was done utilizing the built-in error estimation capabilities within the BEEF-vdW exchange correlation functional, which bounds the barriers obtained using other generalized gradient approximation (GGA)-level exchange correlation functionals (Supplementary Fig. 7).[33] This analysis indicates that the surface of LiF could be highly conductive if surface Li$^+$ ions aren't pinned to any position on the LiF lattice, e.g., by coordinating (i.e., Lewis basic) moieties on the polymer. Notably, unlike conventional polymer electrolytes, such as PEO, PIM-1 does not feature a high density of Li$^+$-coordinating motifs and is characteristically rigid, given its ladder-like backbone features no rotating σ bonds. These unique physical properties serve to un-entangle Li$^+$ transport from polymer segmental chain dynamics at polymer–LiF interfaces and thereby ensures low $V_{Li+}$. Nevertheless, the low density of nitrile-based coordinating motifs on PIM-1 is likely to increase, to a degree, the stability of the lithium at the surface, thereby affecting the barrier for Li$^+$ hopping, which can be quantified experimentally.

We experimentally determined the influence of PIM-1 on the Li$^+$ hopping barrier in LiF@PIM-1 composites by evaluating their temperature-dependent ionic conductivity in Li–Li symmetric cells over a temperature range of 25–85 °C (Fig. 3b,c). Our determination of $E_a$ = 0.42 eV is consistent with those for other solid-ion conductors, and are distinguished from solvent-mediated ion transport, e.g., within the mesopores of polyolefin separators, where $E_a$ = 0.11 eV. Thus, PIM-1 increases the energetic barrier for interfacial Li$^+$ along LiF by only 80 meV, consistent with our hypothesis that the low density of benzonitriles on PIM-1 is beneficial in keeping interfacial Li$^+$ transport pathways minimally obstructed. Further support for interfacial ion transport in LiF@PIM composites is evidenced by the loading-dependent ionic conductivity, which increases by over an order of magnitude ($\sigma_{20C}$ = 2.9–33 µS cm$^{-1}$) for LiF loadings spanning 0.5–8.4% (*w/w*) (Table 1); interestingly, $\sigma_{20C}$ is only 2.2 µS cm$^{-1}$ for PIM-1, indicating the presence of LiF significantly enhances the ionic conductivity of the polymer host, even though the activation barrier increases as PIM-1's gel-polymer electrolyte characteristics fade and the composite's solid-ion conducting characteristics emerge with increased LiF loading. Based on composite morphology, interfacial resistance (Supplementary Fig. 8), and ionic conductivity, we advanced LESA-3 as our optimized formulation for further studies.

To understand the fraction of the ion current contributed by Li$^+$, we determined the cation transference number ($t^+_{ss}$) of LESA-3 during potentiostatic lithium metal plating in Li–Li symmetric cells. The cell was polarized at 10 mV and the current at steady state ($i_{ss}$) was assessed after 10 h (Fig. 3c). We also acquired EIS spectra for the cell before polarization, and at steady-state (Fig. 3d). To calculate $t^+_{ss}$, we employed the Bruce–Vincent method[35]. Notably, LESA-3 exhibits $t^+_{ss}$ = 0.69 ± 0.03, which is higher than that of the carbonate electrolyte ($t^+_{ss}$ ~0.4).[36] This suggests that LESA-3 should limit the extent of space-charge accumulation in the SIC during Li metal plating at high rate.

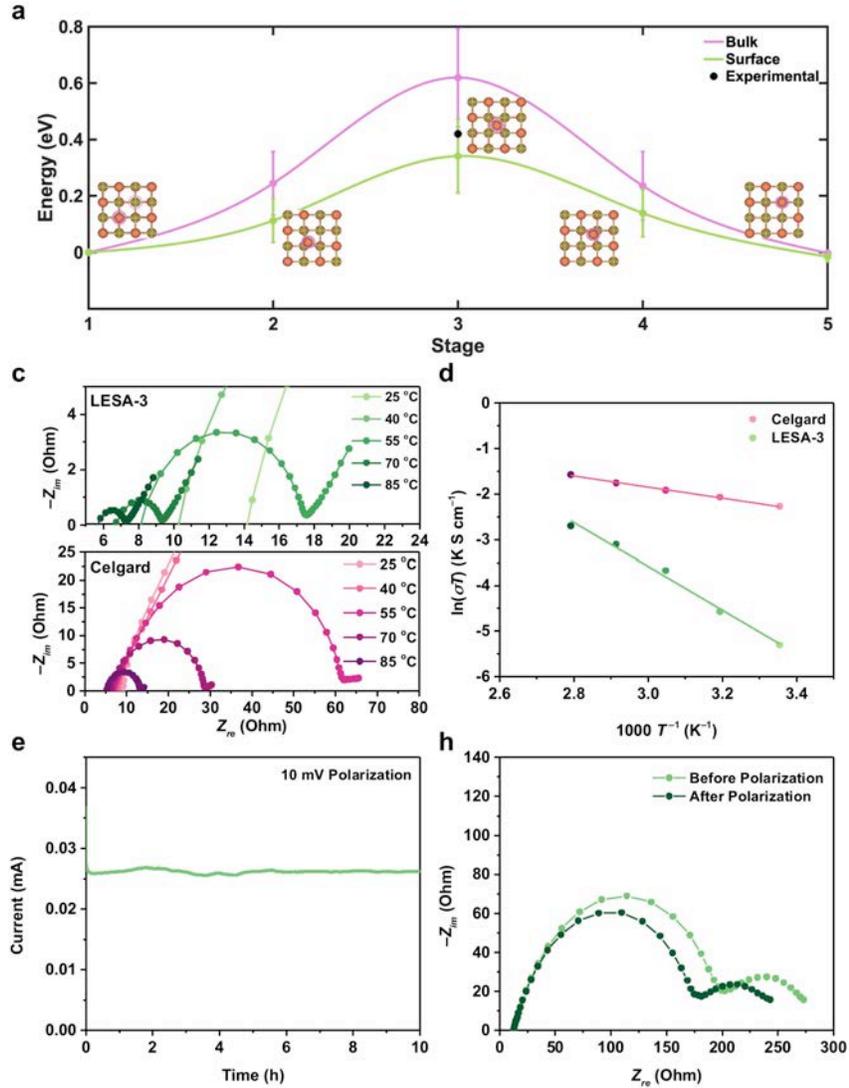

**Fig. 3: Li ion migration in LESA.**
**a,** Energy landscape of Li$^+$ hopping at the surface and within bulk LiF. The insets show the different stages encountered during hopping from a Li site (encircled in solid pink) to a nearby vacant site. The experimentally determined activation barrier for Li$^+$ hopping further takes into account the influence of PIM-1, which results in an 80 meV increase above the calculated result for LiF. **b,** EIS spectra measured at different temperatures for Li–Li symmetric cells assembled with either LiF@PIM (LESA-3) on Celgard 2325 or unmodified Celgard 2325. **c,** Arrhenius plot used to extract activation barriers for ionic conduction within LiF@PIM SICs (LESA-3) and liquid electrolyte-infiltrated Celgard 2325. **d,** Steady-state current determination under 10 mV polarization for Li–Li cells with LiF@PIM SICs (LESA-3). **e,** EIS spectra before and after Li–Li cell polarization.

**Li Metal Dendrite Suppression by LiF@PIM SICs**

To demonstrate the efficacy of LiF@PIM-1 SICs in suppressing dendrite formation during high-rate cycling, we carried out plate–strip tests in symmetric Li–Li cells configured with thin Li electrodes (~30 μm). Cells were cycled at 2 mA cm$^{-2}$ at 20 ˚C with 1 mAh cm$^{-2}$ of the lithium reversibly cycled (~33% of the Li inventory). Those configured with LiF@PIM-1 formulation LESA-3 exhibited superior performance: the initial area-specific resistance (ASR) was ~18 Ω•cm$^2$ and cycle-life was ~270 h, where ~540 mAh cm$^{-2}$ cumulative capacity was cycled (Fig. 4a). In contrast, cells configured with Celgard (negative control) had an initial ASR of 29 Ω•cm$^2$ and only lasted 10 cycles at 2 mA cm$^{-2}$, before manifesting large, irreversible voltage increases; cells configured with PIM-1 on Celgard (positive control) performed in a similar manner, consistent with the predictions of our chemomechanical model (Fig. 4a). EIS spectra (Fig. 4b) were also acquired for cells configured LESA-3 taken before cycling, at steady state, and after cell failure. The initial interfacial resistance ($R_I$ = 135 Ω) and charge-transfer resistance (R$_{CT}$ = 80 Ω) were lower than those for the Celgard-only cell ($R_I$ = 260 Ω; $R_{CT}$ = 150 Ω, see Supplementary Fig. 9). These data corroborate the low ASR observed for LESA-3 cells in Fig. 3a. At steady state, $R_I$ increased slightly, which may be due to the formation of a surface film on Li during cycling, arises from incomplete solvent blocking by the composite. Cell failure was attributed to a soft short, based on the shape of the EIS curve.

These aggressive cycling conditions for cells configured with thin Li electrodes begin to establish the prospects for LiF@PIM-1 composite SICs to advance Li metal battery technology development, but are not direct evidence of dendrite suppression and dense Li plating mediated by the LiF–PIM-1 SICs. To bridge this gap, we carried out synchrotron hard X-ray microtomography[15] to understand the evolution of the advancing and receding Li–SIC interfaces in Li–Li symmetric cells during plating/stripping at 1 mA cm$^{-2}$ for those configured with either LESA-3 or Celgard. Cross-sectional analysis of the X-ray tomographs taken at 0, 1, 4, and 16 h revealed starkly contrasting behavior (Fig. 4c,d). For cells with LESA-3, there was no observable change in Li density throughout the plating experiment, even after 16 h of the polarization. The interface between Li and LESA-3 maintained excellent coherence for >80 μm of Li metal plated. On the other hand, for cells configured with Celgard, pits on the receding lithium electrode were evidenced within 1 h of stripping, and were increasingly prevalent and deep thereafter. Furthermore, after 4 h, the advancing Li electrode showed the onset of low-density lithium deposits, which were apparent across the entire electrode surface; the cell never recovered from this instability. Taken together, these data suggest that LESA-3 should slow the onset of low-density (ramified) Li deposits, and thereby prolong the cycle life of Li metal batteries by slowing electrolyte degradation and mitigating density-driven instabilities that would otherwise lead to dendrite formation and cell shorting.

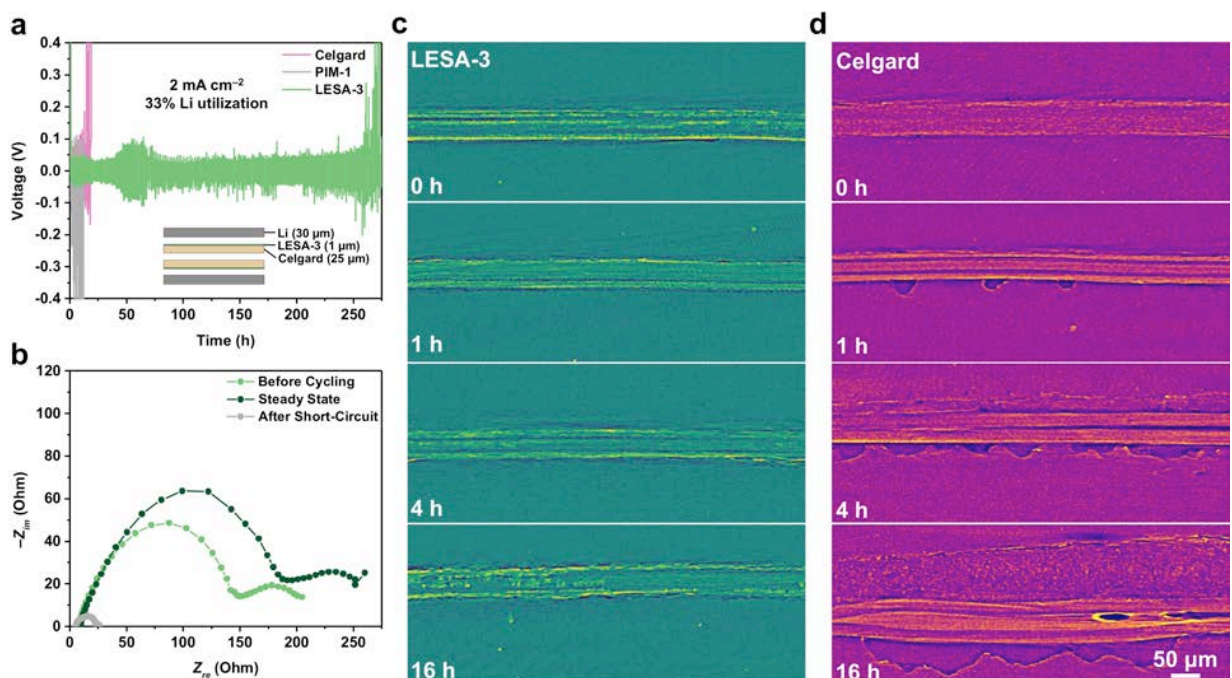

**Fig. 4: Uniform lithium metal electrodeposition enabled by dendrite-suppressing LiF@PIM-1 solid-ion conductors.**
**a,** Galvanostatic cycling of Li–Li cells configured with either LiF@PIM SICs (LESA-3), PIM-1, or Celgard 2325 at 2 mA cm$^{-2}$, where 2 mAh cm$^{-2}$ of Li was (de)plated in each cycle at 20 °C; 30-μm thick Li electrodes were used throughout. **b,** EIS spectra measured for Li–Li cells configured with LiF@PIM SICs (LESA-3) before cycling, during steady-state, and after cycling. **c, d,** Cross-sectional analysis of synchrotron hard x-ray tomograms, visualizing both the advancing and receding electrode–electrolyte interfaces in Li–Li cells configured with either **c** LiF@PIM SICs (LESA-3) or **d** liquid electrolyte-infiltrated Celgard 2325, after the cells had been polarized for different durations at 1 mA cm$^{-2}$.

**Influence of Dendrite-Suppressing LiF@PIM Hybrid SICs on Lithium-Metal Battery Performance**

LESAs comprised of thin lithium anodes laminated with a flexible, adherent, coated separator that interconverts to a thin LiF–PIM-1 dendrite-suppressing solid-ion conductor with low ASR are attractive, drop-in components for lithium metal battery manufacturing. To understand their prospects in that regard, we assembled high-voltage Li–NMC-622 cells with the optimized LESA-3 formulation in place, alongside 30-μm thick Li anodes and 1.44 mAh cm$^{-2}$ NMC-622 cathodes (Fig. 5a). Owing to the low *N/P* ratio, ~20% of the Li inventory was reversibly cycled. During galvanostatic cycling at 1 mA cm$^{-2}$ at 20 °C, the capacity faded at a rate of ~0.07% per cycle while the Coulombic efficiency was >99%. After having reached an initial capacity of ~135 mAh g$^{-1}$, 70% was retained after ~330 cycles. From cycle 1 to cycle 300, a gradual increase in overpotential was observed in both the charge and discharge curves (Fig. 5b), which suggests capacity fade is tied to the increase in cell ASR. Nevertheless, these data compare quite favorably against those for Celgard-only (negative control) and PIM-1-coated Celgard (positive control) (Fig. 5a), where cells reached 70% of their initial capacity after 130 and 100 cycles, respectively. This results in rates of capacity fade of ~0.23% per cycle for Celgard-based cells and ~0.3% per cycle for PIM-1-on-Celgard-based cells.

Despite being a SIC, LiF@PIM-1 composites were unexpectedly more rate-tolerant from 0.1–5 mA cm$^{-2}$ than either Celgard- or PIM-1-on-Celgard-configured Li–NMC-622 cells (Fig. 5c), whose Li-metal plating is liquid-coupled. The divergence was most obvious at 2.0 and 5.0 mA cm$^{-2}$. Furthermore, despite making excursions into high-rate discharge and charge regimes, once the cells returned to cycling at current densities of 1.0 mA cm$^{-2}$, that discharge capacity remained stable at ~130 mAh g$^{-1}$. After 30 cycles, we extracted the lithium anodes from full cells configured with either LESA-3 or Celgard. The anode the LESA-3 cell demonstrated dense Li deposits (Fig. 5d), whereas that for the Celgard cell showed low-density growth of anisotropic Li wires. These data, along with the synchrotron hard X-ray microtomography, continue to highlight how LiF@PIM-1 composites access advantageous dendrite-suppressing character within our chemomechanical paradigm via density-driven architectural design.

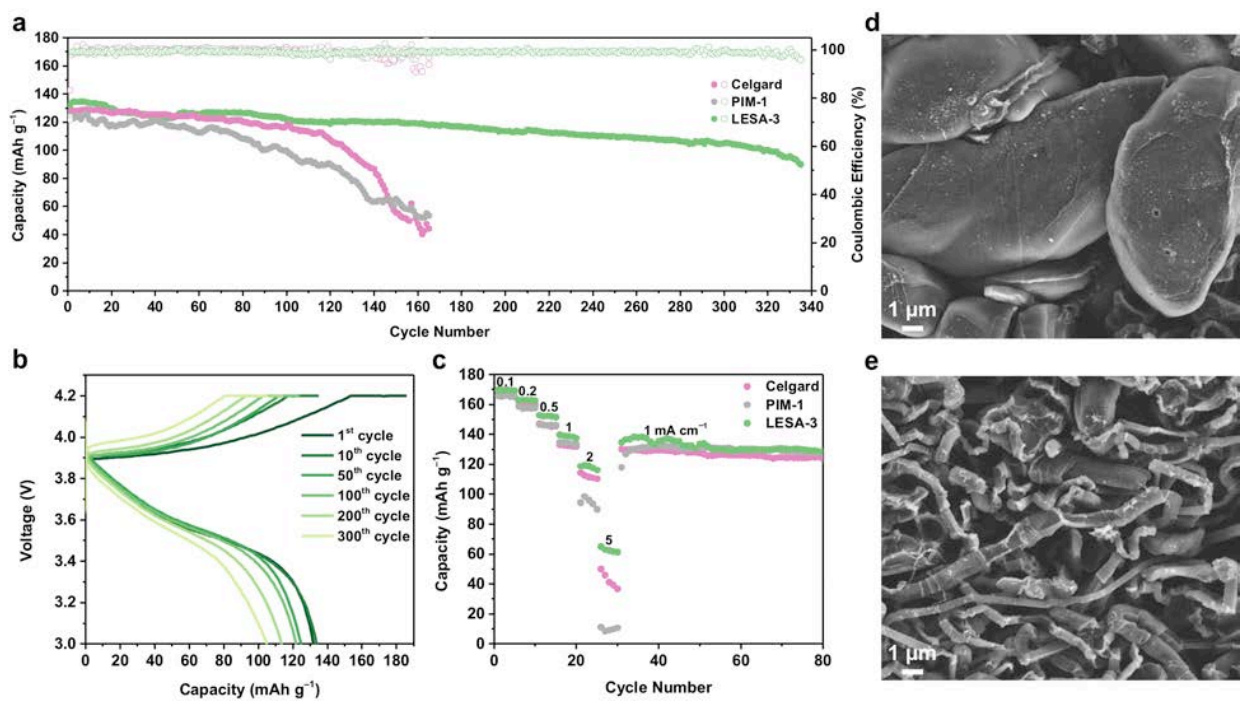

**Fig. 5: Divergent electrochemical performance of Li–NMC-622 cells assembled with thin Li anodes, highlighting the benefits of dendrite-suppressing LiF@PIM solid-ion conductors.** **a,** Galvanostatic cycling at 20 ˚C of Li–NMC-622 cells (30-μm thick Li anode; cathode areal capacity = 1.44 mAh cm$^{-2}$) configured with either LiF@PIM SICs (LESA-3), Celgard (negative control), or PIM-1-coated Celgard (positive control). C-rate was chosen to deliver a current density of 1 mA cm$^{-2}$. **b,** Galvanostatic charge–discharge curves at various stages in the LESA-3 cell's cycling history. **c,** Rate tolerance of Li–NMC-622 cells (30-μm thick Li anode; cathode areal capacity = 1.44 mAh cm$^{-2}$) cycled at 20 ˚C, configured with either LiF@PIM SICs (LESA-3), Celgard (negative control), or PIM-1-coated Celgard (positive control). **d,e,** Top-down SEM images of lithium anodes from the Li–NMC-622 full cells with either **d,** LiF@PIM SICs (LESA-3) or **e,** Celgard after 30 cycles.

**Discussion**

The arc developed here for LiF@PIM composites as dendrite-suppressing SICs in lithium metal batteries, within the larger narrative of a universal chemomechanical model for electrodeposition of metals, indicates that a sensible discourse is now possible when rationalizing both the successes and failures of SICs in preventing dendrite formation, whether the outcome is fundamentally pressure- or density-driven. Furthermore, confining nano-LiF domains to the open-pore network of structurally rigid polymer hosts is an exciting strategy by which to elaborate hybrid SICs that embody the unique transport physics of nanoionics. Macromolecular hosts with tailored interfaces in such hybrids are key to revealing properties and behaviors not typically available to all-inorganic platforms. Here, macromolecular nanoionics directed us to LiF@PIMs as "soft ceramic" SICs due to their unusual chemomechanical characteristics, which are put to work in dendrite-suppression in Li metal batteries. It is likely that similar nanostructured composites will become available using leading-edge sol-gel chemistry,[37] atomic layer deposition,[38] chemical condensation,[39] ball-milling, and other techniques, allowing their use broadly in batteries known to suffer from dendrite-related failure. From such an expanded exploration, it will be further evident that SICs are incapable of accessing both dendrite-suppression and dendrite-blocking character in the same material, as is delineated in our model. That said, we anticipate that there is an end-game, where both can be realized in the cell, as may be necessary for battery safety long term. Specifically, we see soft ceramic ion conductors such as LiF@PIM composites as useful for dendrite suppression, directly at Li metal, where only a thin layer is ultimately needed to take advantage of those characteristics; a second layer, on top of the soft ceramic, can stand-in as the fail-safe dendrite-blocking layer. Layered hybrids of this nature and complexity can be manufactured and integrated into the battery manufacturing infrastructure, suggesting a path forward for commercialization efforts where those batteries are key to widespread electrification of the transportation sector.

**Methods**

**Computational methods.** Self-consistent DFT calculations were carried out using the real-space projector-augmented wave (PAW) method[40,41] as implemented in GPAW[42,43]. To model the LiF surface, slabs were created by repeating a $Li_4F_4$ cell in space to form four layers, two of which (bottom) were fixed. Li atoms were removed from the surface to achieve the desired vacancy concentration. Periodic boundary conditions were used for *x* and *y* directions and a vacuum of 15 Å was used in the *z* direction perpendicular to the surface on both sides of the slab. A real-space grid spacing of 0.18 Å was used and the Brillouin zone was sampled using the Monkhorst Pack scheme[44]. All calculations were converged to energy <0.5 meV and force <0.01 eV Å$^{-1}$. To determine the size of the system that would give the best computational cost-effective accuracy, a study of the effect of periodic images was conducted by investigating the $E_a$ for hopping on surfaces of sizes yielding vacancy concentrations of 4, 5, and 8%. The Perdew–Burke–Ernzerhof (PBE) generalized gradient approximation functional[45] was used for these calculations, and the *k*-point grid chosen was, respectively, *(2×3×1), (3×3×1),* and *(5×3×1)*. The nudged elastic band method as implemented in atomic simulation environment[46] was employed to create five stages for Li hopping on each of these different surfaces. $E_a$ converged within 0.01 eV at a vacancy concentration of 8%, and was used for all subsequent simulations. The Bader charge method[29] was applied at each of the five stages on the 8% vacancy concentration surface to compute the volume of the mobile $Li^+$, which was then used to determine the volume ratio of 0.21 reported in the main body of the paper. To quantify uncertainty in $E_a$, the energies of these five stages obtained earlier were calculated using the BEEF-vdW functional[33], which gives an ensemble of 2000 different energy values for each stage. Of these values, only those that yield a physically meaningful description of the energy landscape were used for the calculation of the average and uncertainty estimation. A similar process was used to determine the bulk LiF hopping barrier: an LiF structure periodic in all directions was created and a Li atom was removed, creating a vacancy concentration of 1.4%. The values for energy barrier for bulk and surface were then compared for each functional. The ratio between conductivity for each system is proportional to the exponential of the difference in hopping barrier divided by $k_BT$ (Supplementary Fig. 7).

**Preparation of coated separators.** Inks of PIM-1[25,47] (100 mg mL$^{-1}$) and TBAF (0–60% *w/w*) in chloroform were blade-coated onto Celgard 2535 to form luminous yellow coatings. Coated separators were dried overnight at 50 °C *in vacuo* before coin cell assembly.[48]

**Materials characterization.** SEM images were acquired using a Zeiss Gemini Ultra-55 analytical SEM at beam energy of 3 keV. For XRD and TGA, inks comprised of PIM-1 and TBAF were drop-cast on glass slides and dried overnight at 50 °C *in vacuo*. Cation metathesis was carried out using 1.0 M $LiPF_6$ in EC:DMC (1:1) in Ar-glovebox for 48 h, after which the samples were washed with DMC and dried overnight at 50 °C *in vacuo*. XRD was conducted using a Bruker AXS D8 Discover GADDS X-Ray diffractometer (XRD), which is operated at 35 kV and 40 mA with a Co Kα wavelength of 1.79Å. TGA was conducted using a TA Instruments Q5500 and performed in Ar from room temperature to 600 °C with a heating ramp of 10 °C min$^{-1}$, and a 2-h isothermal step at 600 °C. For DMA, freestanding films were cast in 3-cm diameter glass wells, and the solvent allowed to evaporate over 24 h. The films were then incubated in 1.0 M $LiPF_6$ in EC:DMC (1:1) in an Ar-glovebox for 72 h to generate the composites, which were then incubated in DMC overnight (4 times) and dried overnight at 50 °C *in vacuo*. Dynamic mechanical analysis (DMA) was conducted using a TA Instruments Q800, where stress−strain measurements were performed at room temperature up to 15 N with a stress ramp of 0.5 N min$^{-1}$.

**Electrochemical characterization.** Electrochemical studies were performed using CR2032 coin cells. A given LiF@PIM formulation, LESAs 1–6, or PIM-1 was coated on Celgard 2325 and assembled such that the coating was in direct contact with the lithium-metal electrode surface. Li–Li cells were assembled with thin lithium (30 μm) on copper foil (10 μm); the liquid electrolyte was 1.0 M $LiPF_6$ in EC:DMC (1:1). Li–NMC-622 cells were assembled with 30-μm thick Li anodes and NMC-622 ($LiNi_{0.6}Mn_{0.2}Co_{0.2}O_2$) cathodes with areal capacity of 1.44 mAh cm$^{-2}$, which were provided by the CAMP facility at Argonne National Lab; the liquid electrolyte was 1.0 M $LiPF_6$ in EC:DMC (1:1) with 10% *w/w* fluoroethylene carbonate (FEC)[49] and 1% *w/w* vinylene carbonate (VC)[50].

**Synchrotron hard X-ray microtomography.** Polarized Li–Li cells were dissembled, the Li/electrolyte/Li sandwiches were punched to quarter inch disks, and the disks were sealed in Al-laminated pouches in an Ar-glovebox. Monochromatric hard X-ray (23 keV) microtomography was then carried out on beamline 8.3.2 at the Advanced Light Source at Lawrence Berkeley National Laboratory.[15] The samples were rotated 180° under the X-ray, and the shadows cast by the samples were converted to image stacks with ~1,500 images in each stack. The stacks were re-sliced with Tomviz software to obtain the cross-sectional tomography slices.

**Data availability**

The data that support the findings of this study are available from the corresponding authors upon reasonable request.

**References**

1. Xu, W., Wang, J., Ding, F., Chen, X., Nasybulin, E., Zhang, Y. & Zhang, J.-G. Lithium metal anodes for rechargeable batteries. *Energy Environ. Sci.* **7**, 513–537 (2014).
2. Tikekar, M. D., Choudhury, S., Tu, Z. & Archer, L. A. Design principles for electrolytes and interfaces for stable lithium-metal batteries. *Nat. Energy* **1**, 16114 (2016).
3. Lin, D., Liu, Y. & Cui, Y. Reviving the lithium metal anode for high-energy batteries. *Nat. Nanotech.* **12**, 194–206 (2017).
4. Cheng, X.-B., Zhang, R., Zhao, C.-Z. & Zhang, Q. Toward safe lithium metal anode in rechargeable batteries: a review. *Chem. Rev.* **117**, 10403–10473 (2017).
5. Manthiram, A., Yu, X. & Wang, S. Lithium battery chemistries enabled by solid-state electrolytes. *Nat. Rev. Mater.* **2**, 16103 (2017).
6. Albertus, P., Babinec, S., Litzelman, S. & Newman, A. Status and challenges in enabling the lithium metal electrode for high-energy and low-cost rechargeable batteries. *Nat. Energy* **3**, 16–21 (2018).
7. Monroe, C. & Newman, J. Dendrite growth in lithium/polymer systems. *J. Electrochem. Soc.* **150**, A1377–A1384 (2003).
8. Monroe, C. & Newman, J. The effect of interfacial deformation on electrodeposition kinetics. *J. Electrochem. Soc.* **151**, A880–A886 (2004).
9. Ahmad, Z. & Viswanathan, V. Stability of electrodeposition at solid-solid interfaces and implications for metal anodes. *Phys. Rev. Lett.* **119**, 056003 (2017).
10. Zachman, M. J., Tu, Z., Choudhury, S., Archer, L. A. & Kourkoutis, L. F. Cryo-STEM mapping of solid–liquid interfaces and dendrites in lithium-metal batteries. *Nature* **560**, 345–349 (2018).
11. Barai, P., Higa, K. & Srinivasan V. Lithium dendrite growth mechanisms in polymer electrolytes and prevention strategies. *Phys. Chem. Chem. Phys.* **19**, 20493–20505 (2017).
12. Singh, M. *et al.* Effect of molecular weight on the mechanical and electrical properties of block copolymer electrolytes. *Macromolecules* **40**, 4578-4585 (2007).
13. Bouchet, R. *et al.* Single-ion BAB triblock copolymers as highly efficient electrolytes for lithium-metal batteries. *Nat. Mater.* **12**, 452–457 (2013).
14. Khurana, R., Schaefer, J. L., Archer, L. A. & Coates, G. W. Suppression of lithium dendrite growth using cross-linked polyethylene/poly(ethylene oxide) electrolytes: a new approach for practical lithium-metal polymer batteries. *J. Am. Chem. Soc.* **136**, 7395–7402 (2014).
15. Harry, K. J., Hallinan, D. T., Parkinson, D. Y., MacDowell, A. A. & Balsara, N. P. Detection of subsurface structures underneath dendrites formed on cycled lithium metal electrodes. *Nat. Mater.* **13**, 69–73 (2014).
16. Porz, L. *et al.* Mechanism of lithium metal penetration through inorganic solid electrolytes. *Adv. Energ. Mater.* **7**, 1701003 (2017).

17. Cheng, E. J., Sharafi, A. & Sakamoto, J. Intergranular Li metal propagation through polycrystalline $Li_{6.25}Al_{0.25}La_3Zr_2O_{12}$ ceramic electrolyte. *Electrochim. Acta.* **223**, 85–91 (2017).
18. Maier, J. Nanoionics: ion transport and electrochemical storage in confined systems. *Nat. Mater.* **4**, 805–815 (2005).
29. Liang, C. C. Conduction characteristics of the lithium iodide–aluminum oxide solid electrolytes. *J. Electrochem. Soc.* **120**, 1289–1292 (1973).
20. Weinstein, L., Yourey, W., Gural, J. & Amatucci, G. Electrochemical Impedance Spectroscopy of Electrochemically Self-Assembled Lithium–Iodine Batteries. *J. Electrochem. Soc.* **155**, A590–A598 (2008).
21. Wu, F. *et al.* Lithium iodide as a promising electrolyte additive for lithium–sulfur batteries: mechanisms of performance enhancement. *Adv. Mater.* **27**, 101–108 (2015).
22. Lu, Y., Tu, Z. & Archer, L. A. Stable lithium electrodeposition in liquid and nanoporous solid electrolytes. *Nat. Mater.* **13**, 961–969 (2014).
23. Fan, L., Zhuang, H. L., Gao, L., Lu, Y. & Archer, L. A. Regulating Li deposition at artificial solid electrolyte interphases. *J. Mater. Chem. A* **5**, 3483–3492 (2017).
24. Li, H., Richter, G. & Maier, J. Reversible formation and decomposition of LiF clusters using transition metal fluorides as precursors and their application in rechargeable Li batteries. *Adv. Mater.* **15**, 736–739 (2003).
25. Budd, P. M. *et al.* Solution-processed, organophilic membrane derived from a polymer of intrinsic microporosity. *Adv. Mater.* **16**, 456–459 (2004).
26. Li, C. *et al.* Engineered transport in microporous materials and membranes for clean energy technologies. *Adv. Mater.* **30**, 1704953 (2018).
27. Myung, S.-T. *et al.* Nickel-rich layered cathode materials for automotive lithium-ion batteries: achievements and perspectives. *ACS Energy Lett.* **2**, 196–223 (2017).
28. Newman, J. & Chapman, T. W. Restricted diffusion in binary solutions. *AlChE J.* **19**, 343–348 (1973).
29. Bader, R. F. W. Atoms in Molecules: A Quantum Theory (Oxford Univ. Press, New York, 1990).
30. Bader, R. F. W., Carroll, M. T., Cheeseman, J. R. & Chang, C. Properties of atoms in molecules: atomic volumes. *J. Am. Chem. Soc.* **109**, 7968–7979 (1987).
31. Ozhabes, Y., Gunceler, D. & Arias, T. A. Stability and surface diffusion at lithium–electrolyte interphases with connections to dendrite suppression. Preprint at http://arxiv.org/abs/1504.05799 (2015).
32. Sharada, S. M., Bligaard, T., Luntz, A. C., Kroes, G.-J., & Nørskov, J. K. SBH10: A benchmark database of barrier heights on transition metal surfaces. *J. Phys. Chem. C* **121**, 19807–19815 (2017).
33. Wellendorff, J. *et al.* Density functionals for surface science: Exchange-correlation model development with Bayesian error estimation. *Phys. Rev. B* **85**, 235149 (2012).
34. Berne, B. J., Cicotti, G. & Coker, D. F. Classical and quantum dynamics in condensed phase simulations. (World Scientific, Villa Marigola, 1998).
35. Bruce, P. G., Evans, J. & Vincent, C. A. Conductivity and transference number measurements on polymer electrolytes. *Solid State Ionics* **28–30**, 918–922 (1988).
36. Suo, L., Hu, Y.-S., Li, H., Armand, M. & Chen, L. A new class of Solvent-in-Salt electrolyte for high-energy rechargeable metallic lithium batteries. *Nat. Comm.* **4**, 1481 (2013).

37. Orilall, M. C. & Wiesner, U. Block copolymer based composition and morphology control in nanostructured hybrid materials for energy conversion and storage: solar cells, batteries, and fuel cells. *Chem. Soc. Rev.* **40**, 520–535 (2011).
38. Peng, Q., Tseng, Y.-C., Darling, S. B. & Elam, J. W. Nanoscopic patterned materials with tunable dimensions via atomic layer deposition on block copolymers. *Adv. Mater.* **22**, 5129–5133 (2010).
39. Llordes, A. *et al.* Linear topology in amorphous metal oxide electrochromic networks obtained via low-temperature solution processing. *Nat. Mater.* **15**, 1267–1273 (2016).
40. Blochl, P. E. Projector augmented-wave method. *Phys. Rev. B* **50**, 17953 (1994).
41. Kresse, G. & Joubert, D. From ultrasoft pseudopotentials to the projector augmented-wave method. *Phys. Rev. B* **59**, 1758 (1999).
42. Mortensen, J. J., Hansen, L. B. & Jacobsen, K. W. Real-space grid implementation of the projector augmented wave method. *Phys. Rev. B* **71**, 035109 (2005).
43. Enkovaara, J. *et al.* Electronic structure calculations with GPAW: a real-space implementation of the projector augmented-wave method. *J. Phys. Condens. Matter* **22**, 253202 (2010).
44. Monkhorst, H. J. & Pack, J. D. Special points for Brillouin-zone integrations. *Phys. Rev. B* **13**, 5188 (1976).
45. Perdew, J. P., Burke, K. & Ernzerhof, M. Generalized Gradient Approximation Made Simple. *Phys. Rev. Lett.* **77**, 3865 (1996).
46. Larsen, A. H. *et al.* The Atomic Simulation Environment–A Python library for working with atoms. *J. Phys. Condens. Matter* **29**, 273002 (2017).
47. Li, C. *et al.* A Polysulfide-Blocking Microporous Polymer Membrane Tailored for Hybrid Li–Sulfur Flow Batteries. *Nano Lett.* **15**, 5724–5729 (2015).
48. Ward, A. L. *et al.* Materials genomics screens for adaptive ion transport behavior by redox-switchable microporous polymer membranes in lithium–sulfur batteries. *ACS Cent. Sci.* **3**, 399–406 (2017).
49. Markevich, E., Salitra, G. & Aurbach, D. Fluoroethylene carbonates an important component for the formation of an effective solid electrolyte interphase on anodes and cathodes for advanced Li-ion batteries. *ACS Energy Lett.* **2**, 1337–1345 (2017).
50. Aurbach, D., Gamolsky, K., Markovsky, B., Gofer, Y., Schmidt, M. & Heider, U. On the use of vinylene carbonate(VC) as an additive to electrolyte solutions for Li-ion batteries. *Electrochim. Acta* **47**, 1423–1439 (2002).

**Author contributions**

B.A.H. designed and directed the study. C.F. and A.W.E. conducted the experiments. V.V. and Z.A. conducted the simulations. V.V. designed and directed the theoretical study. B.A.H. and C.F. wrote the paper with contributions from all co-authors.

**Competing interests**

B.A.H. is an inventor on PCT patent application 62/431,300 submitted by Lawrence Berkeley National Laboratory that covers these and related classes of solid-ion conductors, as well as aspects of their use in electrochemical devices.

**Acknowledgments**

This work was supported by the Advanced Research Projects Agency-Energy Integration and Optimization of Novel Ion Conducting Solids (IONICS) program under Grant No. DE-AR0000774. Z.A. was supported in part by the Phillips and Huang Family Fellowship in Energy from the College of Engineering at Carnegie Mellon University. A.W.E. was supported by the U.S. Department of Energy, Office of Science, Office of Workforce Development for Teachers and Scientists (WDTS) under the Science Undergraduate Laboratory Internship (SULI) program. Portions of this work, including polymer synthesis and characterization, were carried out as a User Project at the Molecular Foundry, which is supported by the Office of Science, Office of Basic Energy Sciences, of the U.S. Department of Energy under Contract No. DE-AC02-05CH11231. Synchrotron hard x-ray tomography was conducted on BL 8.3.2 at the Advanced Light Source, which is a DOE Office of Science User Facility under the same contract. NMC-622 cathodes were provided by Bryant J. Polzin from the Cell Analysis, Modeling, and Prototyping (CAMP) Facility at Argonne National Laboratories. The computational portion of this work was performed on the Hercules computer cluster, which was funded through a Carnegie Mellon College of Engineering Equipment grant. Y. Wang, D. Prendergast, D. Parkinson, P. Frischmann, Y.-M. Chiang, P. Albertus, S. Babinec, and D. Cagle are thanked for helpful discussions.

**Additional information**

Supplementary information is available for this paper at XXX.

Reprints and permissions information are available at www.nature.com/reprints.

Correspondence and requests for materials should be addressed to B.A.H.

**Supporting Information**

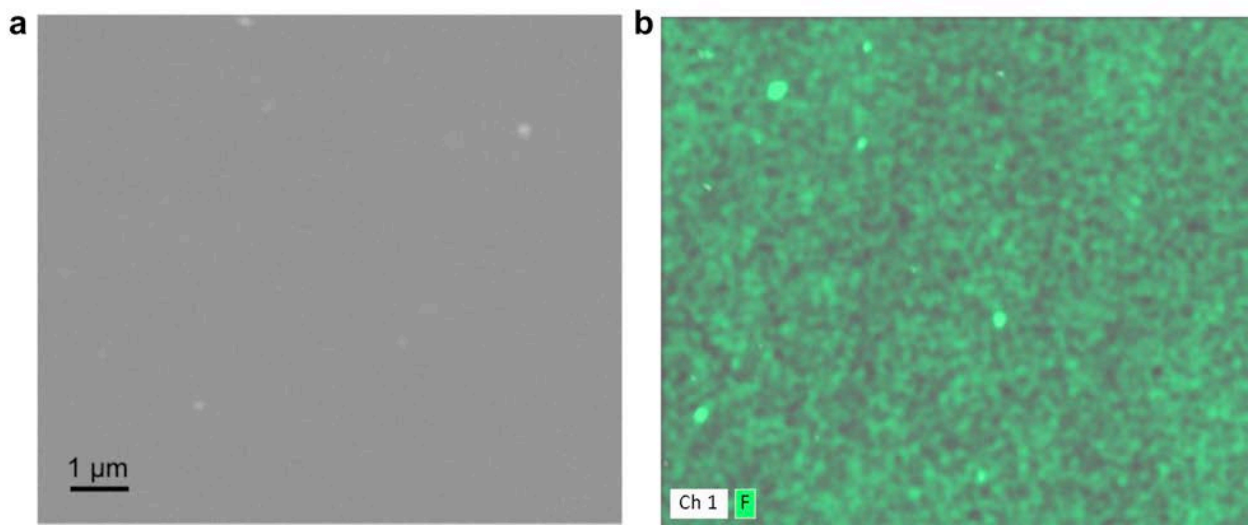

**Supplementary Fig. 1 | Top-Down scanning electron micrograph (SEM) of LESA-3 and EDS mapping of elemental fluorine distribution. a,** Lateral smoothness of LiF@PIM composite formulation LESA-3 was evident in the SEM. **b,** EDS map in the imaged area showed uniform distribution of F, indicating homogeneous distribution of LiF nanostructured domains within the composite SIC.

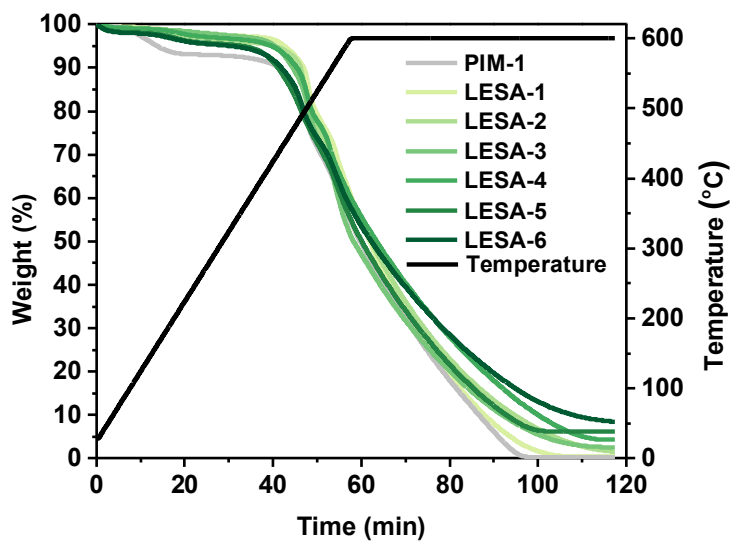

**Supplementary Fig. 2 | Thermogravimetric determination of LiF loading in LiF@PIM composite SICs generated by cation metathesis.** TGA was performed in Ar from room temperature to 600 °C with a heating ramp of 10 °C min$^{-1}$, and a 1-h isothermal step at 600 °C. PIM-1 decomposes with no char residue, therefore the residual mass for LESAs 1–6 can be attributed to LiF. By varying the TBAF loading in PIM-1, the resulting LiF loading after cation exchange can be tuned, which was demonstrated here over the range of 0.5–8.4% *w/w*.

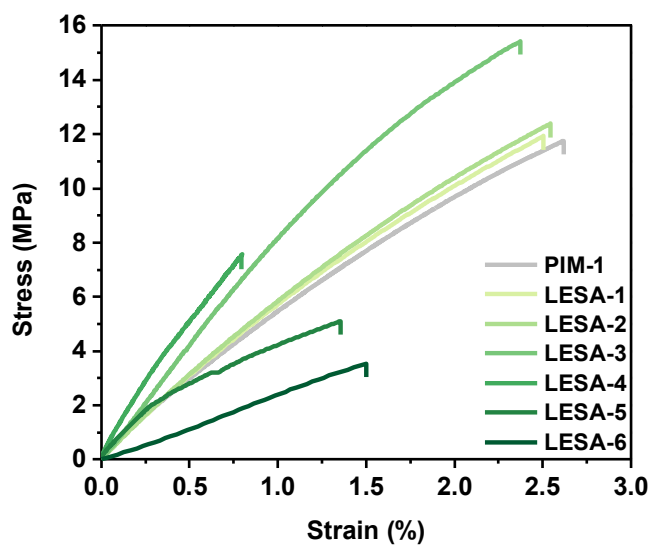

**Supplementary Fig. 3 | Stress–Strain curves for PIM-1 and LESAs 1–6.** Dynamic mechanical analysis (DMA) was conducted using a TA Instruments Q800, where stress–strain measurements were performed at room temperature up to 15 N with a stress ramp of 0.5 N min$^{-1}$. Assuming a Poisson's ratio of 0.33, PIM-1 has a shear modulus, $G_S$, of 165 MPa. At relatively low LiF content, $G_S$ of LiF@PIM composites increase with LiF loading. $G_S$ reaches a maximum value of 376 MPa when the LiF loading is 4.3% w/w. When LiF content is increased further, $G_S$ decreases significantly, which is linked to the emergence of porosity due to the associated volume changes inherent to the cation metathesis. Composites with higher LiF loading experience less deformation before breaking, indicating their brittleness.

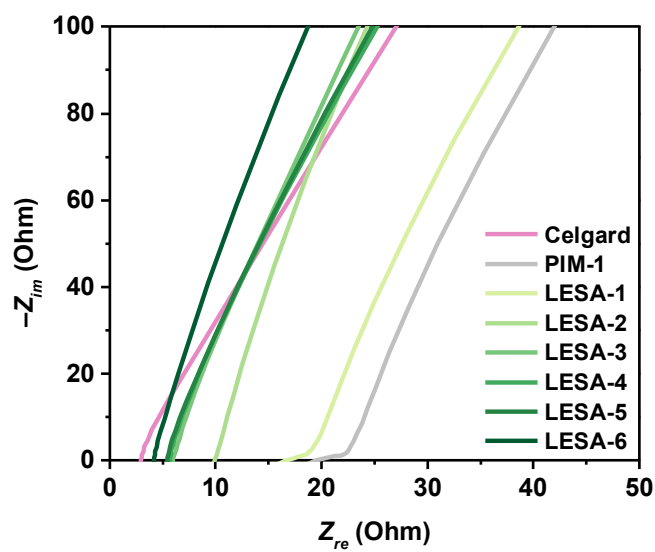

**Supplementary Fig. 4 | EIS spectra of symmetric stainless steel cells incorporating LiF@PIM SICs with variable LiF loading.** Stainless steel spacers were used as the electrodes, between which was sandwiched a single layer of LiF@PIM composite, LESAs 1–6. The high-frequency intercept of each cell's EIS spectrum at 20 °C was used to calculate the SIC's ionic conductivity, $\sigma_{20C}$ (Table 1).

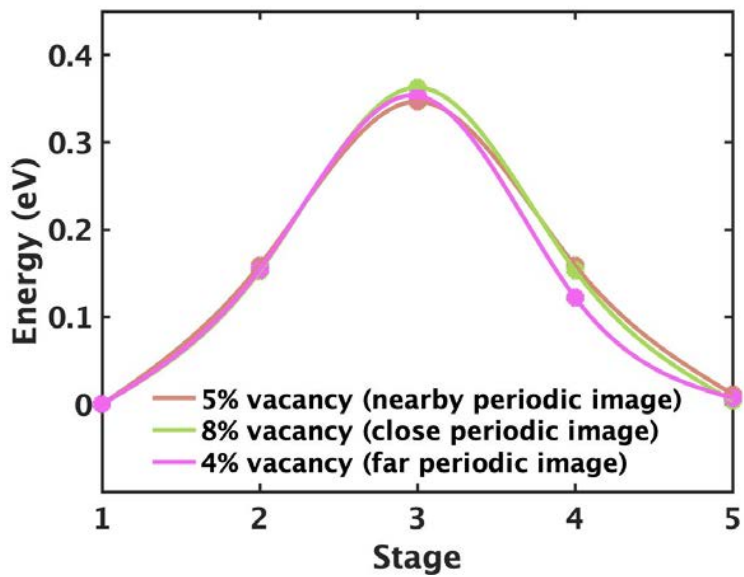

**Supplementary Fig. 5 | Periodic image effects on activation energy of hopping on LiF surface.** The NEB method is used to calculate the activation barrier along with the required number of supercells to achieve the desired Li vacancy concentrations of 4, 5, and 8%. The analysis shows that the activation energy is converged to within 10 meV at a Li vacancy concentration of 8%.

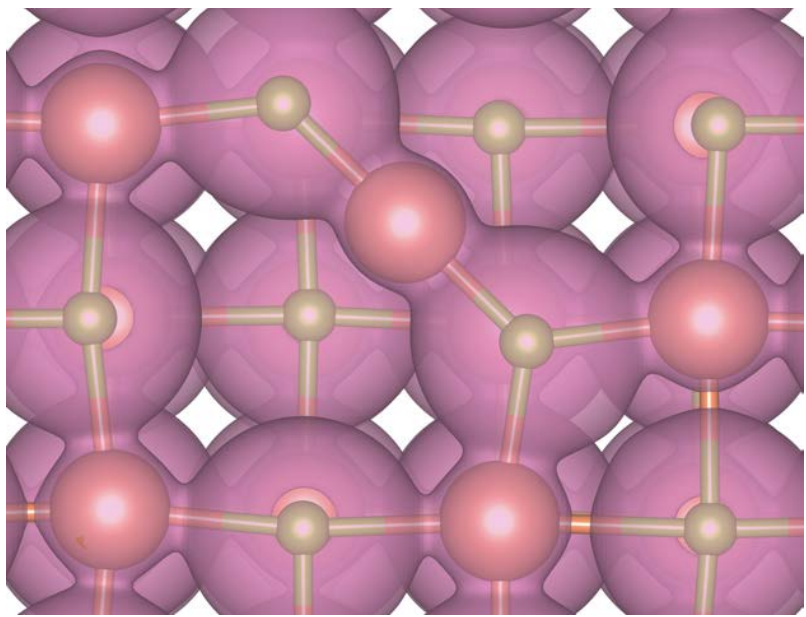

**Supplementary Fig. 6 | Isosurfaces of charge density on LiF surface during Li hopping.** The charge density is obtained using self-consistent DFT. The Bader volume for each atom is calculated by partitioning the density into zero-flux surfaces.

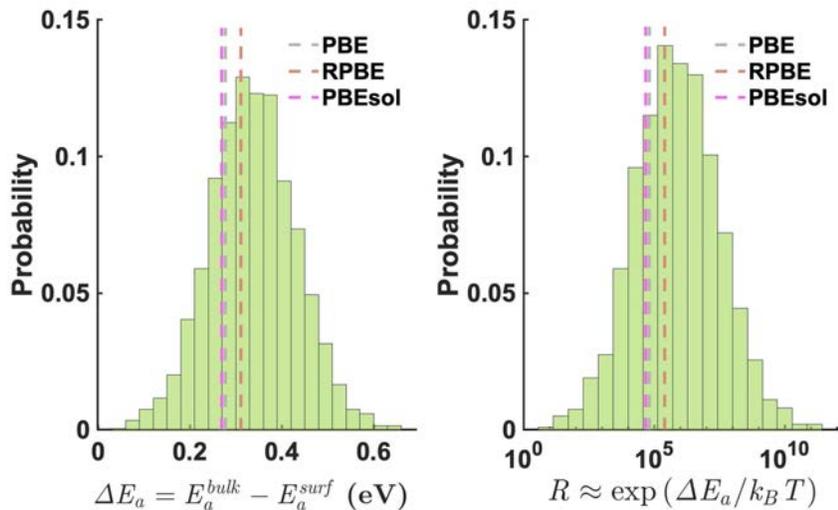

**Supplementary Fig. 7 | Probability distribution of the difference in activation energy of Li hopping between bulk and surface of LiF.** The difference $\Delta E_a = E_a^{bulk} - E_a^{surf}$ is shown in a histogram corresponding to the calculation performed using an ensemble of exchange-correlation functionals. The ratio of ionic conductivity $R \approx \exp(\Delta E_a/k_B T)$ in each system at 300 K is also shown as a histogram. The values for specific exchange correlation functionals (PBE[45], revised PBE, and PBEsol) are marked with dashed lines. The figure shows bulk ionic conduction in LiF is insignificant compared to surface.

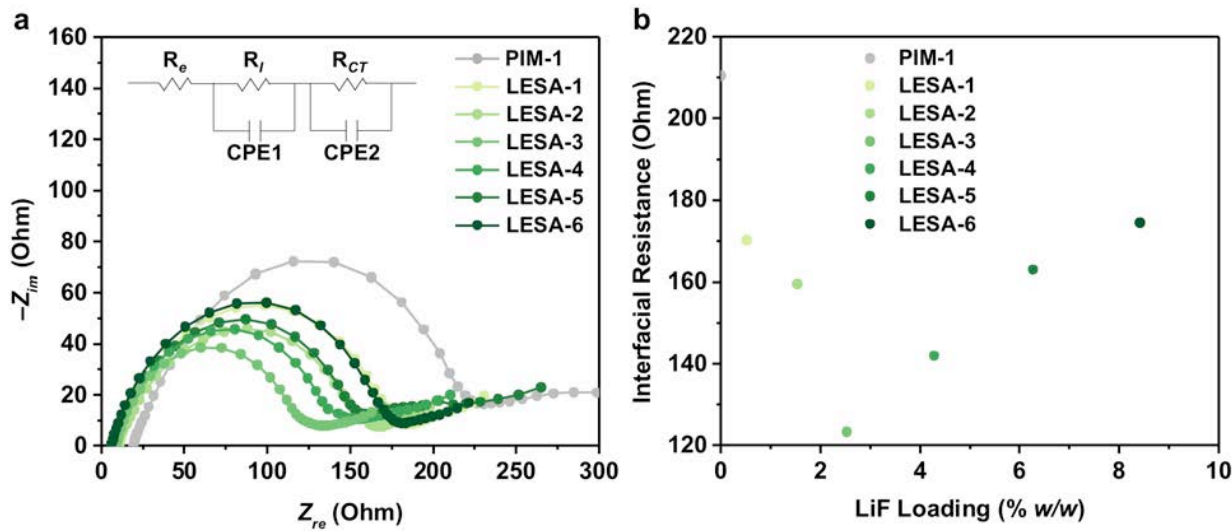

**Supplementary Fig. 8 | EIS spectra and interfacial resistance of Li–Li cells incorporating PIM-1 and LESAs 1–6, which vary in LiF loading. a,** EIS spectra measured at 20 °C for Li–Li cells incorporating LESAs 1–6 and PIM-1. **b,** Interfacial resistance of the cells extracted from the EIS spectra. All EIS curves show two semicircles: the first real-axis intercept of the curves represents the electrolyte resistance, $R_e$, the first semicircle is related to the electrode–electrolyte interfacial resistance, $R_I$, and the second semicircle is related to the charge–transfer resistance, $R_{CT}$. The interfacial resistance, $R_I$, was plotted as a function of the composite's LiF content.

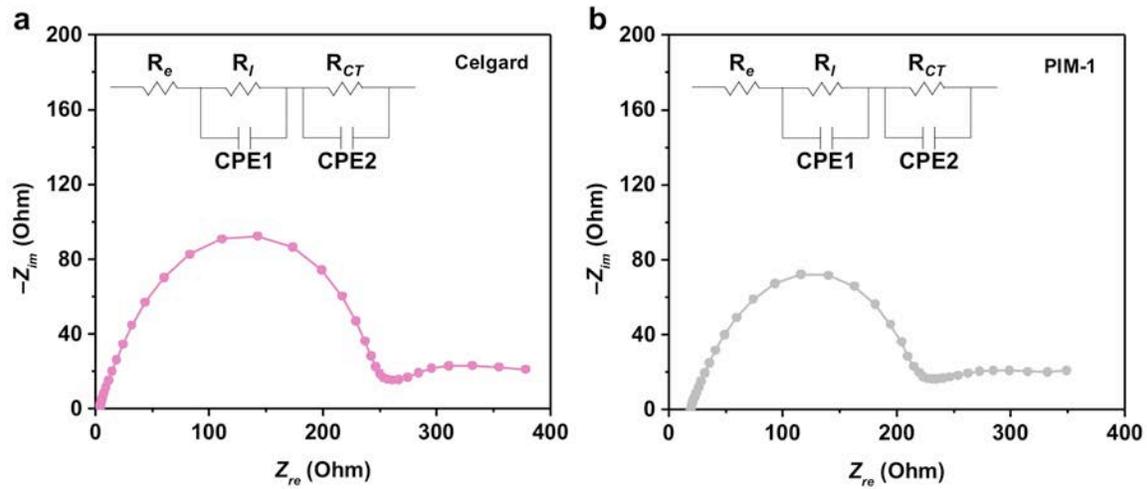

**Supplementary Fig. 9 | EIS spectra of Li–Li cells. a**, EIS for Li–Li cells assembled with unmodified Celgard 2325 separators (negative control). **b,** EIS for Li–Li cells assembled with PIM-1 coated Celgard 2325 separators (positive control). The EIS curves show two semicircles: the first real-axis intercept of the curves represents the electrolyte resistance, $R_e$, the first semicircle is related to the electrode–electrolyte interfacial resistance, $R_I$, and the second semicircle is related to the charge–transfer resistance, $R_{CT}$. Both controls show significantly higher $R_I$ than those configured with LESAs 1–6 (Supplementary Fig. 8a).